\pdfoutput=1

\documentclass[twocolumn,showpacs]{revtex4}
\usepackage{graphicx,amssymb,amsmath}
\begin{document}

\title{
  Strain-induced transitions to quantum chaos and effective time-reversal
  symmetry breaking in triangular graphene nanoflakes
}

\author{Adam Rycerz}
\affiliation{Marian Smoluchowski Institute of Physics, 
Jagiellonian University, Reymonta 4, PL--30059 Krak\'{o}w, Poland}

\begin{abstract}
We investigate the effect of strain-induced gauge fields on statistical distribution of energy levels of triangular graphene nanoflakes with zigzag edges. In the absence of strain fields but in the presence of weak potential disorder such systems were found in Ref.\ \cite{Ryc12} to display the spectral statistics of the Gaussian unitary ensemble (GUE) due to the effective time-reversal ({\em symplectic}) symmetry breaking. Here show that, in the absence of disorder,  strain fields may solely lead to spectral fluctuations of GUE providing a~nanoflake is deformed such that all its geometric symmetries are broken. In a~particular case when a~single mirror symmetry is preserved the spectral statistics follow the Gaussian orthogonal ensemble (GOE) rather then GUE. The corresponding transitions 
to quantum chaos are rationalized by means of additive random-matrix models and the analogy between strain-induced gauge fields and real magnetic fields is discussed.
\end{abstract}

\date{\today}
\maketitle

\section{Introduction}
Graphene, a~two-dimensional form of carbon, provides an intriguing condensed-matter analogue of $(2\!+\!1)$ dimensional quantum electrodynamics \cite{Sem84}. Several unique features of graphene, not found in conventional metals, semimetals, and insulators, follow from the effective description for low-energy excitations given by the Dirac equation for spin-$1/2$ fermions with zero rest mass \cite{Cas09,Gei09,Das11}. Unlike a massless neutrino described by a~similar Weyl equation \cite{Abe04}, each effective quasiparticle in graphene has the electric charge $-e$ and is coupled to external electromagnetic fields via scalar and vector potential terms \cite{Bee08}, offering several ways to tune the quantum states in order to control graphene-based electronic devices. Additionally, the dynamics of carriers in graphene can be affected by mechanical deformations inducing effective gauge fields \cite{Cas09,Suz02,Voz10}. As strains exceeding $10\%$ can be applied to graphene in a~reversible manner \cite{Lee08}, this new way of controlling the electronic structure has attracted significant theoretical \cite{Fog08,Per09,Gui10a,Cho10,Gui10b,Jua11} and experimental \cite{Boo08,Bun08,Bao09,Hua09,Moh09,Lev10} attention. Remarkably, the Landau quantization signalling the presence of strain-induced pseudomagnetic fields greater than $300\,$T was recently demonstrated \cite{Lev10}. For the opposite limit of weak strain fields, the existence of a~zero magnetic field analogue of the Aharonov-Bohm effect in graphene \cite{Sch12} is predicted theoretically \cite{Jua11}.

Surprisingly, the influence of strain fields on quantum chaotic behavior of electrons confined in graphene quantum dots \cite{Pon08} has not been discussed so far. Quantum chaotic behavior appears generically for systems, whose classical dynamics are chaotic, and manifests itself via the fact that energy levels show statistical fluctuations following those of Gaussian ensembles of random matrices \cite{Haa10}. In particular, if such a~system posses the time-reversal symmetry (TRS), its spectral statistics follow the Gaussian orthogonal ensemble (GOE). A~system with TRS and half-integer spin has the symplectic symmetry and, in turn, shows spectral fluctuations of the Gaussian symplectic ensemble (GSE). If TRS is broken, as in the presence of nontrivial gauge fields, and the system has no other antiunitary symmetry \cite{Rob86}, spectral statistics follow the Gaussian unitary ensemble (GUE). For a particular case of massless spin-$1/2$ particles, it was pointed out by Berry and Mondragon \cite{Ber87}, that the confinement may break TRS in a~persistent manner (i.e., even in the absence of gauge fields), leading to the spectral fluctuations of GUE \cite{XNi12}.

When applying the above symmetry classification to graphene nanosystems \cite{Ryc12,Wur09} one needs, however, to take into account that Dirac fermions in graphene appear in the two valleys, $K$ and $K'$, coupled by TRS \cite{valefoo}. If the valley pseudospin is conserved, a~special (symplectic) time-reversal symmetry (STRS) becomes relevant, playing a~role of an effective TRS in a~single valley \cite{Wur09}. Both real magnetic and strain-induced gauge fields may break STRS leading to the spectral fluctuations of GUE. As demonstrated numerically in Ref.\ \cite{Ryc12}, such fluctuations also appear for particular {\em closed} nanosystems in graphene in the presence of random scalar potentials slowly varying on the scale of atomic separation. Such nanosystems include equilateral triangles with zigzag or Klein edges, i.e., with terminal atoms belonging to one sublattice. Generic graphene nanoflakes with irregular edges show spectral fluctuations of GOE \cite{Wur09,Lib09}, as strong intervalley scattering restores TRS in the absence of gauge fields. In contrast, the boundary effects are suppressed in {\em open} graphene systems, for which signatures of the symplectic symmetry class were reported \cite{Tik08}.

\begin{figure}
\centerline{\includegraphics[width=0.9\linewidth]{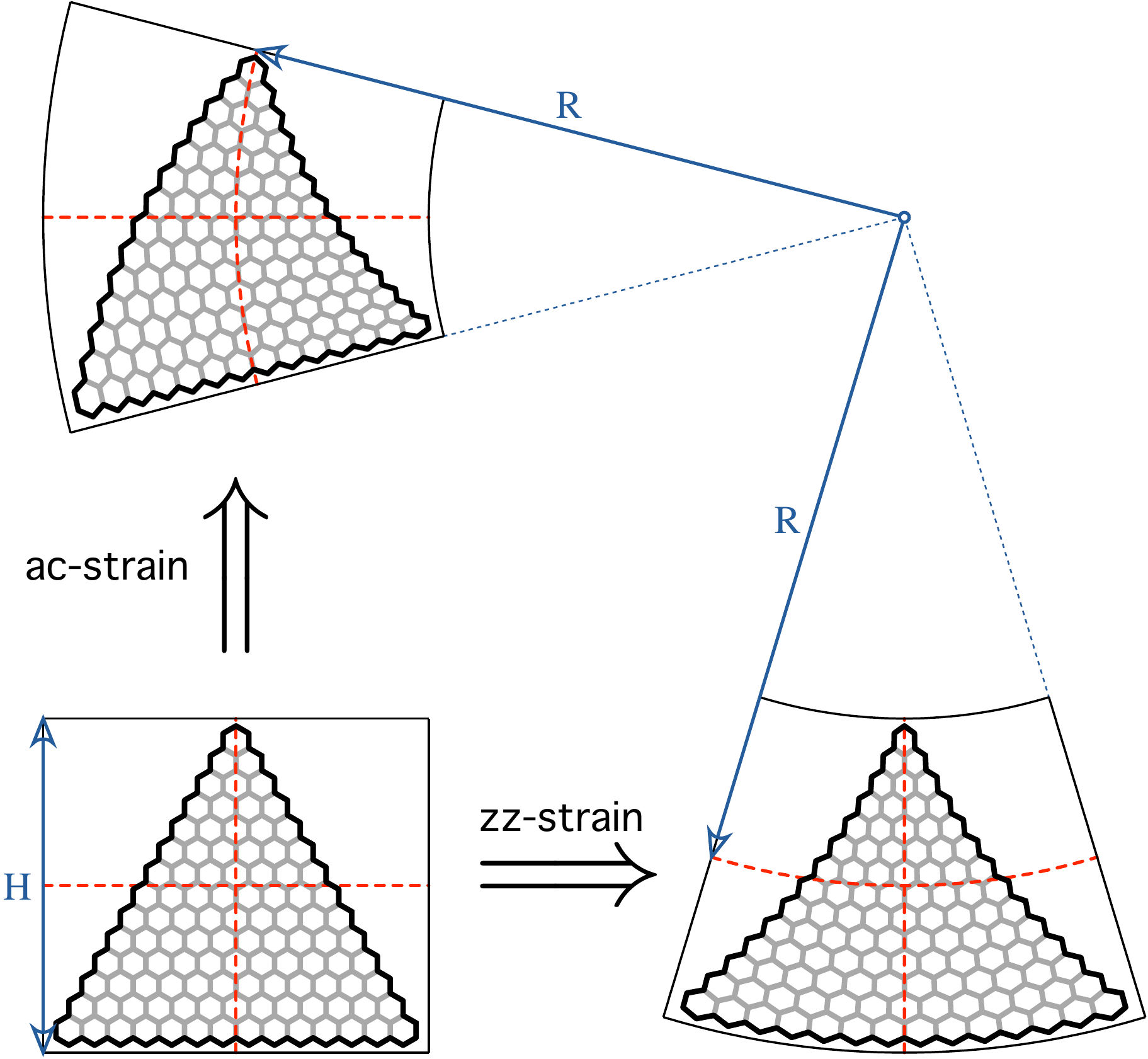}}
\caption{\label{strageo}
  Systems studied numerically in the paper. Bottom left: Triangular graphene nanoflake with zigzag edges characterized by the height $H$. Remaining plots: The same system bent in-plain employing the strain geometry proposed by Guinea {\em et al.}\ \cite{Gui10b} (top left) in the variant breaking all geometric symmetries ({\em ac-strain}), and (bottom right) in the variant preserving the mirror symmetry ({\em zz-strain}). The radii of arcs limiting the flake area are $R\pm{}H/\sqrt{3}$ for ac-strain or $R\pm{}H/2$ for zz-strain. The ratio $H/R=2$ in both cases. 
} 
\end{figure}

In this paper, we analyze numerically the spectral statistics of triangular graphene nanoflakes with zigzag edges bent in-plane according to the strain geometry proposed by Guinea {\em at al.}\ \cite{Gui10b} (see Fig.\ \ref{strageo}). In the absence of strain fields and the disorder, such systems were found in Ref.\ \cite{Ryc12} to show the Poissonian distribution of energy levels, which gradually evolves towards GUE when increasing the disorder strength indicating the transition to quantum chaos. Although the experimental energy resolution seems not sufficient as yet to discuss spectral statistics, regular graphene nanoflakes, including triangular ones bounded entirely with zigzag edges, were recently fabricated on metallic substrates \cite{Ham11}. Also, strain-induced gauge fields were shown to affect quantum states of similar nanosystems \cite{Lev10}. Here, we demonstrate numerically, that experimentally realistic strains may lead to clear signatures of quantum chaos even in a nanosystem which is otherwise perfect, i.e., not subjected to substrate-induced disorder, atomic-scale defects, etc. We also discuss the role of a~particular strain geometry applied, which may lead to the {\em true} or {\em false} STRS breaking, in a~similar way as the geometry of chaotic Schr\"{o}dinger systems leads to the true or false TRS breaking in the presence of magnetic fields \cite{Rob86}.

The paper in organized as follows. In Sec.\ \ref{strainga}, we briefly discuss how magnetic fields and geometric strains contribute to the effective Dirac Hamiltonian for low-energy spectrum of graphene. In Sec.\ \ref{specsta}, we present our main results concerning the statistical distribution of energy levels of triangular graphene nanoflakes with zigzag edges. A~quantitative comparison of the effect of uniform magnetic field and the effects of different strain-induced fields on such a distribution is given in Sec.\ \ref{reremag}. The conclusions are given in Section \ref{conclu}.

\begin{figure}
\centerline{\includegraphics[width=0.9\linewidth]{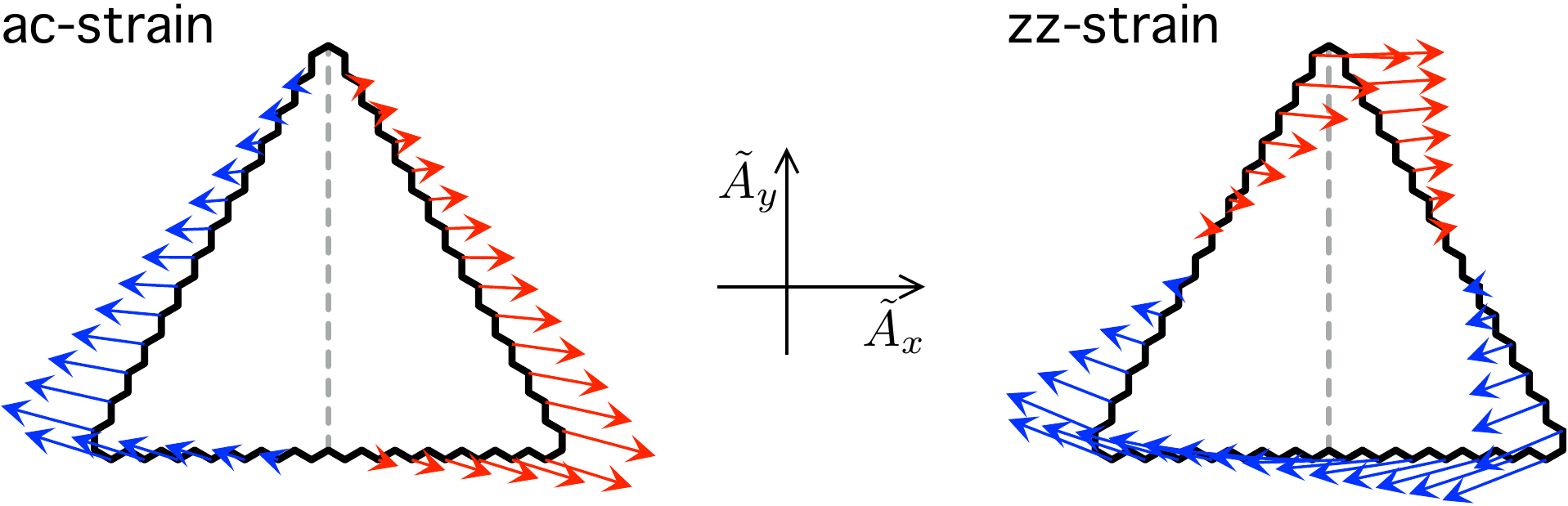}}
\caption{\label{weakstr}
  Strain-induced gauge fields appearing on the edges of triangular nanoflake for in-plane deformations of Fig.\ \ref{strageo} in the limit of $R\gg{}H$ (schematic). Red [blue] arrows are for $\tilde{A}_x>0$ [$\tilde{A}_x<0$]. 
Left: ac-strain leaves only to the approximate $S_x$ symmetry, which is broken by a~small $\tilde{A}_y$ term; see Eq.\ (\ref{aaprop}). Right: zz-strain breaks both ${\cal T}_{\rm sl}$ and $S_x$, but it leaves the exact invariance under the reflection $\mbox{\boldmath$\tilde{A}$}\rightarrow{}-\!\mbox{\boldmath$\tilde{A}$}$ combined with $S_x$, corresponding to the antiunitary symmetry ${\cal T}_0S_x$ ({\em false} symplectic time-reversal symmetry breaking). 
}
\end{figure}

\section{ Strain-induced gauge fields and Dirac fermions in~graphene 
  \label{strainga} }
We start from the tight-binding Hamiltonian, including the nearest-neighbor hopping-matrix elements between $\pi$ orbitals on a~honeycomb lattice \cite{Cas09,dilafoo}
\begin{multline}
  \label{hamtba}
  {\cal H}_{\rm TB}= -t_0\sum_{\langle{}ij\rangle}
  \left(1-\beta\frac{\delta{}d_{ij}}{d_0}\right) \\
  \times 
  \left[ \exp\left(i\frac{2\pi}{\Phi_0}\int_i^j{\bf A}\cdot{}d{\bf l}\right)
    |i\rangle\langle{}j| + {\rm h.c.} \right],
\end{multline}
where $\delta{}d_{ij}/d_0$ is the relative change in bond length, with $d_0=a/\sqrt{3}$ being the equilibrium bond length defined via the lattice spacing in graphene $a=0.246\,$nm, and ${\bf A}$ is the vector potential related to the real magnetic field by ${\bf B}=\mbox{rot}{\bf A}$ and incorporated as a Peierls phase (with the flux quantum $\Phi_0=h/e$). Remaining parameters are the equilibrium hopping integral $t_0\simeq{}3\,$eV and the dimensionless electron-phonon coupling $\beta=-\partial\log(t)/\partial\log(d)|_{d=d_0}\simeq{}2-3$ \cite{Voz10}.

\subsection{The effective Hamiltonian and its symmetries}
It can be shown, that for low-energy excitations, low magnetic fields, and deformations slowly varying on the scale of atomic separation $a$, ${\cal H}_{\rm TB}$ (\ref{hamtba}) reduces to the effective Dirac Hamiltonian \cite{Sem84,Suz02,Voz10}
\begin{equation}
  \label{hameff}
  {\cal H}_{\rm eff}= 
  v_F\tau_0\mbox{\boldmath$\sigma$}\cdot
  \left(\mbox{\boldmath$p$}\!+\!e{\bf A}\right) 
  - v_F\tau_z\mbox{\boldmath$\sigma$}\cdot\mbox{\boldmath$\tilde{A}$},
\end{equation}
where $v_F=3t_0d_0/(2\hbar)\simeq{}10^6\,$m/s is the energy-independent Fermi velocity, $\mbox{\boldmath$\sigma$}=(\sigma_x,\sigma_y)$, $\sigma_i$ and $\tau_i$ ($i=1,2,3$) are the Pauli matrices acting on sublattice and valley degrees of freedom (respectively), $\sigma_0$ ($\tau_0$) denotes the unit matrix, and $\mbox{\boldmath$p$}=-i\hbar(\partial_x,\partial_y)$ is the in-plane momentum operator. ${\cal H}_{\rm eff}$ (\ref{hameff}) is written in the so-called valley-isotropic representation \cite{Bee08}; namely, it acts on spinors $\psi\equiv[\psi_A,\psi_B,-\psi_B',-\psi_A']^T$, where $A/B$ is the the sublattice index, the primed and unprimed entries correspond to $K$ and $K'$ valleys. The strain-induced gauge fields can be written as 
\begin{equation}
  \label{tildeaa}
  \mbox{\boldmath$\tilde{A}$}\equiv 
  \left(\begin{array}{c}  \tilde{A}_x \\ \tilde{A}_y \end{array}\right)= 
  \frac{c\beta}{d_0}
  \left(\begin{array}{c} u_{xx}-u_{yy} \\ -2u_{xy} \end{array}\right),
\end{equation}
where $c$ is a dimensionless coefficient of the order of unity, $u_{ij}=\frac{1}{2}(\partial_iu_j+\partial_ju_i)$ ($i,j=1,2$) is the strain tensor for in-plane deformations \cite{Lan59}, and we have chosen the coordinate system $(x,y)$ such that the $x$ axis corresponds to a zigzag direction of a~honeycomb lattice.

In the absence of gauge fields (${\bf A}=\mbox{\boldmath$\tilde{A}$}=0$), the Hamiltonian (\ref{hameff}) is invariant upon the antiunitary operations ${\cal T}_k^{-1}{\cal H}_{\rm eff}{\cal T}_k$ with \cite{Wur11}
\begin{equation}
  {\cal T}_k= \sigma_y\otimes\tau_k{\cal C},\ \ \ \ k=0,1,2,
\end{equation}
where ${\cal C}$ denotes complex conjugation. We notice that ${\cal T}_y^2=1$, and thus ${\cal T}_y$ represents the true time reversal coupling the two valleys; whereas ${\cal T}_0^2={\cal T}_x^2=-1$, leading to the Kramer's degeneracy of the two valleys and to the additional Kramer's degeneracy in each valley. It is easy to see that real magnetic fields ${\bf A}\neq{}0$ break all the symmetries associated with ${\cal T}_k$-s. To the contrary, introducing the strain fields $\mbox{\boldmath$\tilde{A}$}\neq{}0$ and keeping ${\bf A}=0$, one may only break the invariance under the symplectic time reversal ${\cal T}_0$ (STRS), leaving TRS and the invariance under the valley exchange ${\cal T}_x$ unaffected \cite{edgefoo}. 

The above discussion is complete, providing we ignore noncollinear local magnetization (which may appear on the edges of a graphene nanoflake \cite{Sep10,Pot12}), so one can assume that the boundary condition to the effective Dirac equation ${\cal H}_{\rm eff}\psi=E\psi$ preserves TRS \cite{Bee08}. Recent first-principle study by Potasz {\em et al.}\/ \cite{Pot12} shows, that local magnetization decays relatively fast when small graphene flake is charged out of the neutrality point. Although the extrapolation of this result onto much larger flakes as considered here may require some further analysis, it seems natural to expect, that the magnetization does not affect statistical properties of energy levels significantly \cite{magnfoo}. Such an assumption if further supported by the fact, that no disambiguous effects of edge magnetization were observed experimentally even for the lowest-lying energy levels of graphene nanoflakes so far. For these reasons, the effects of possible edge magnetization are neglected in the remaining parts of the paper.

\subsection{Deformations considered in the paper}
We further limit our discussion to the deformations earlier considered in Ref.\ \cite{Gui10b}, namely
\begin{equation} 
  \label{acstra}
  \left(\begin{array}{c} u_x \\ u_y \end{array}\right)_{\rm ac} =
  \left[\begin{array}{c}
      (x-R)\cos\theta_{\rm ac}(y)+R \\
      (R-x)\sin\theta_{\rm ac}(y)
    \end{array}\right]
\end{equation}
with $\theta_{\rm ac}(y)=\left(2y/H\right)\arcsin\left[H/(2R)\right]$, and 
\begin{equation}
  \label{zzstra}
  \left(\begin{array}{c} u_x \\ u_y \end{array}\right)_{\rm zz} =
  \left[\begin{array}{c}
      (R-y)\sin\theta_{\rm zz}(x) \\
      (y-R)\cos\theta_{\rm zz}(x)+R
    \end{array}\right]
\end{equation}
with $\theta_{\rm zz}(x)=\left(\sqrt{3}\,x/H\right)\arcsin\left[H/(\sqrt{3}\,R)\right]$. $R$ is the bending radius of the deformation applied to equilateral triangle of the height $H$; see Fig.\ \ref{strageo}. The labels 'ac' and 'zz' indicate that the strain following from  Eq.\ (\ref{acstra}), hereinafter referred as {\em ac-strain}, predominantly affects the bonds oriented along an armchair direction, whereas the strain following from  Eq.\ (\ref{zzstra}) ({\em zz-strain}) predominantly affects the bonds oriented along a~zigzag direction of a~honeycomb lattice. The maximal strain in the triangle area is 
\begin{equation}
  \label{maxdelij}
  \mbox{max}\left(\frac{\delta{}d_{ij}}{d_0}\right) \simeq 
  \frac{H}{R}\times
  \begin{cases}
    \frac{1}{3}\sqrt{3} & \text{for ac-strain}, \\
    \frac{1}{2} & \text{for zz-strain}.
  \end{cases}
\end{equation}

We notice, that ac-strain breaks all geometric symmetries of the triangular nanoflake for any $H/R>0$, while zz-strain preserves the symmetry with respect to a~mirror reflection $S_x:\ (x,y)\rightarrow(-x,y)$. Such symmetries of the deformations map onto the symmetries of effective gauge fields $\mbox{\boldmath$\tilde{A}$}$ in the Hamiltonian (\ref{hameff}). Keeping only the terms of the order of $R^{-1}$ and $R^{-2}$ one can express the strain fields following from Eqs.\ (\ref{acstra}) and (\ref{zzstra}) for $R\gtrsim{}H$ as
\begin{equation}
  \label{aaprop}
  \mbox{\boldmath$\tilde{A}$}_{\rm ac}\propto
  \left(\begin{array}{c} R^{-1}x \\ R^{-2}xy \end{array}\right)
  \ \ \text{and}\ \ 
  \mbox{\boldmath$\tilde{A}$}_{\rm zz}\propto
  \left(\begin{array}{c} -R^{-1}y \\ R^{-2}xy \end{array}\right).
\end{equation}
In Fig.\ \ref{weakstr}, we plot $\mbox{\boldmath$\tilde{A}$}_{\rm ac}$ and $\mbox{\boldmath$\tilde{A}$}_{\rm zz}$ on the flake edges only, as the fields inside the flake smoothly interpolates between extreme values which are reached at the system boundaries. It is clear that ac-strain breaks the symmetry with respect to $S_x$ (depicted with dashed vertical lines) due to the term $(\tilde{A}_y)_{\rm ac}\propto{}R^{-2}$. For zz-strain, $S_x$ symmetry is also broken by the term $(\tilde{A}_x)_{\rm zz}\propto{}R^{-1}$. Both strain distributions break STRS, as they are not invariant under the transformation $\mbox{\boldmath$\tilde{A}$}\rightarrow -\mbox{\boldmath$\tilde{A}$}$. However,  $\mbox{\boldmath$\tilde{A}$}_{\rm zz}$ exhibits the symmetry under the antiunitary operation ${\cal T}_0S_x$ (with $({\cal T}_0S_x)^2=-1$). In the absence of intervalley scattering and in the limit of quantum chaos, such a~symmetry shall lead to false STRS breaking and spectral fluctuations following GOE rather then GUE \cite{Ber87,Rob86}.

\section{ Spectral statistics \label{specsta} }
This section presents the central results of the paper, concerning statistical distribution of energy levels for Dirac fermions confined in graphene nanoflakes, in the presence of gauge fields introduced in Sec.\ \ref{strainga}. For the numerical illustration, we took an equilateral triangle with zigzag edges containing $N_{\rm C}=32758$ carbon atoms, corresponding to $H=270\,d_0$ and the physical sample area of ${\cal A}_{\rm S}\simeq{}(29\,\mbox{nm})^2$. Such a~system was previously found to show negligibly weak intervalley scattering in the absence of magnetic field, as well as spectral fluctuations following GUE in the presence of disorder smoothly varying on the length scale of $a$ (see Ref.\ \cite{Ryc12}).

\begin{figure}
\centerline{\includegraphics[width=\linewidth]{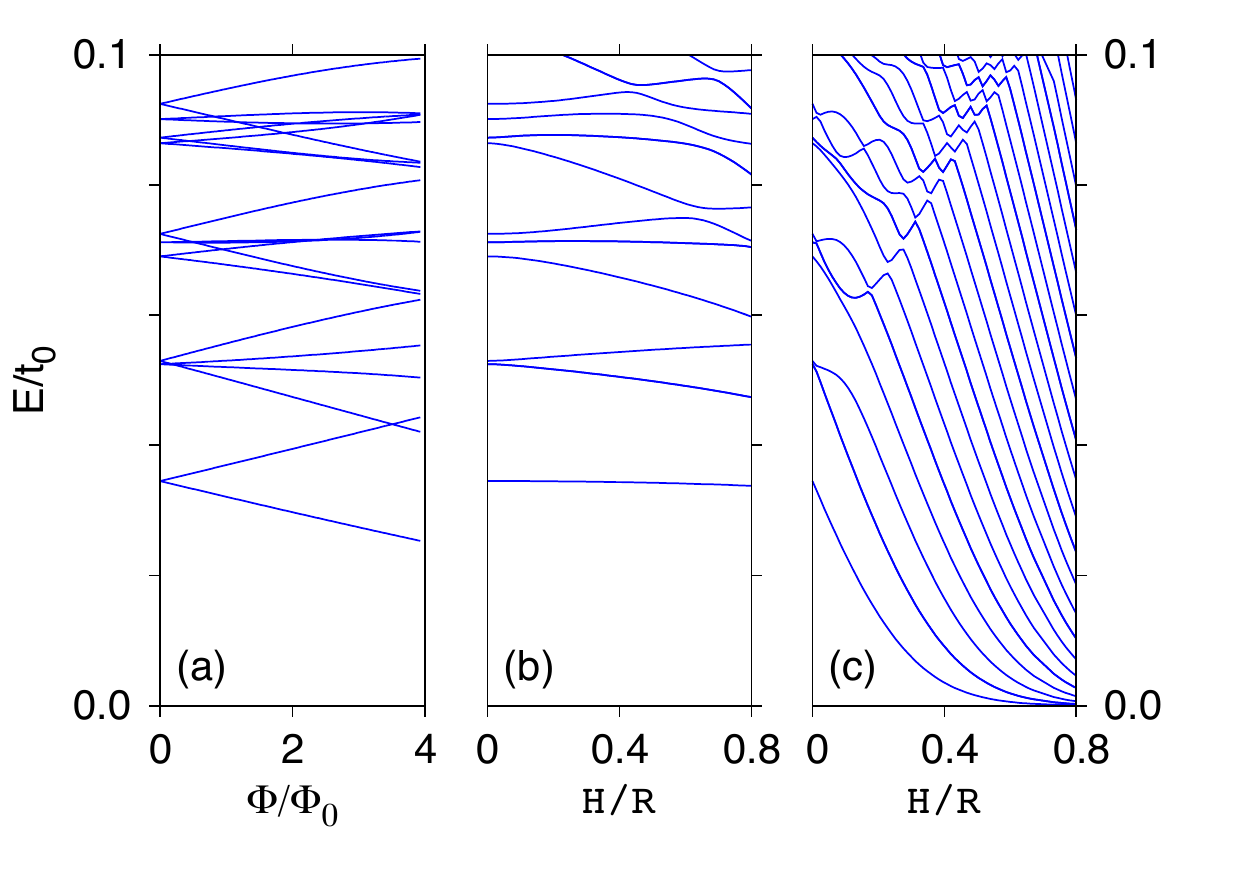}}
\caption{\label{levrep}
  Evolution of energy levels for a~triangular nanoflake containing $N_{\rm C}=32758$ carbon atoms with varying magnetic field (a), ac-strain (b), and zz-strain (c).
}
\end{figure}

\subsection{The effects of magnetic and strain fields}
In Fig.\ \ref{levrep}, we plot energy levels of our model system obtained by numerical diagonalization of ${\cal H}_{\rm TB}$ (\ref{hamtba}) as functions of the magnetic flux $\Phi={\cal A}_{\rm S}B$ [Fig.\ \ref{levrep}(a)], and the parameter $H/R$ characterizing ac-strain or zz-strain [Figs.\ \ref{levrep}(b) or \ref{levrep}(c)]. We limit ourselves to extreme cases of uniform magnetic field applied in the absence of geometric deformations, and two distinct deformations given by Eqs.\ (\ref{acstra}) and (\ref{zzstra}) in the absence of magnetic fields, as intermediate situations simply combine the features of these extreme cases. First, it is clear from Fig.\ \ref{levrep}(a), that uniform magnetic field splits the valley degeneracy of electronic levels \cite{Rec07} and leads to level crossings characteristic for integrable systems. To the contrary, both ac-strain and zz-strain preserve the valley degeneracy and lead to avoided crossings [see Figs.\ \ref{levrep}(b) and \ref{levrep}(c)] characteristic for chaotic quantum systems \cite{Haa10}. A secondary difference between the effects of ac-strain and zz-strain on the electronic structure of the system considered is related to the fact, that different deformations lead to different pseudomagnetic fields. Namely, Eq.\ (\ref{aaprop}) leads to
\begin{equation}
  \label{bsprop}
  B_{{\rm s},\xi}=\hat{\bf e}_z\cdot\mbox{rot}\mbox{\boldmath$\tilde{A}$}_{\xi}\propto
  \frac{H}{R}\left(\eta_{\xi}+\frac{y}{R}\right),
\end{equation}
with $\eta_\xi=0$ if $\xi={\rm ac}$ or $\eta_\xi=1$ if $\xi={\rm zz}$. It is clear from Eq.\ (\ref{bsprop}), that  the strain-induced Landau quantization may be observed in small systems \cite{delnfoo} for zz-strain only (see also Ref.\ \cite{Gui10b}). As this issue is beyond the scope of the paper, we only notice that the systematic drift of lowest-lying electronic levels towards the zero-energy Landau level is totally absent in Fig.\ \ref{levrep}(b) and visible in Fig.\ \ref{levrep}(c). For this reason, the arrangement of ac-strain seems particularly interesting when studying strained graphene nanosystems. In principle, ac-strain may allow one to discuss physical effects of strain-induced gauge fields in the absence of pseudomagnetic fields, in analogy to the Aharonov-Bohm effect appearing for real gauge fields in the absence of magnetic fields, providing the system geometry is modified appropriately \cite{Sch12}.

\begin{figure}
\centerline{\includegraphics[width=\linewidth]{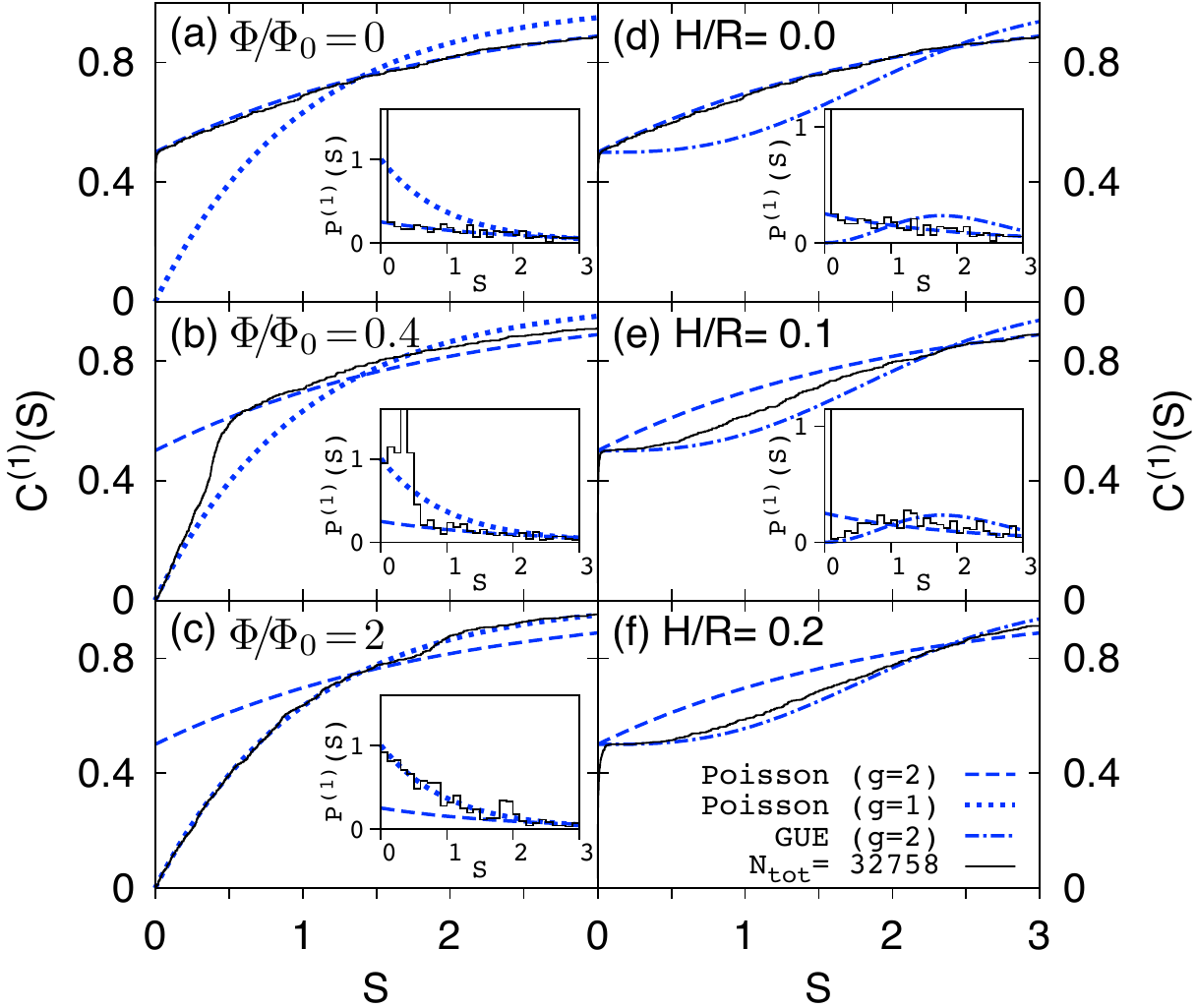}}
\caption{\label{csfirh}
  Integrated level-spacing distributions $C^{(1)}(S)$ for the same system as in Fig.\ \ref{levrep}, in the presence of uniform magnetic field (a)--(c) or ac-strain (d)--(f). The values of total flux $\Phi$ or the strain parameter $H/R$ are specified for each panel. Insets show nearest-neighbor spacings distributions $P^{(1)}(S)$. Numerical results are shown with black solid lines. Remaining lines are for Poisson distribution with the degeneracy $g=2$ of each level [blue dashed], with $g=1$ [blue dotted], and for the Wigner surmise for GUE [blue dash-dotted].
}
\end{figure}

\begin{figure}
\centerline{\includegraphics[width=\linewidth]{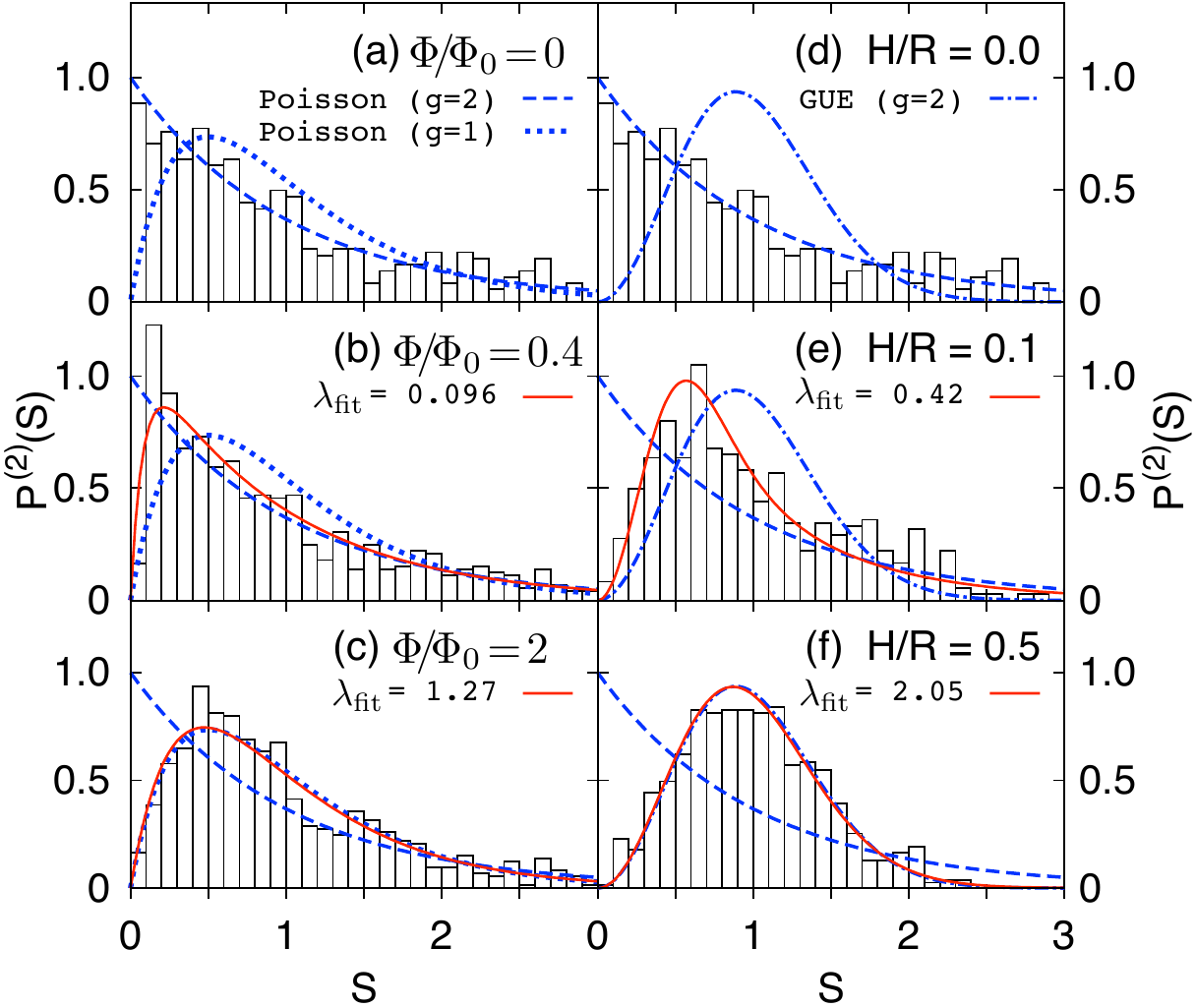}}
\caption{\label{psfirh}
  Second-neighbor level-spacing distributions $P^{(2)}(S)$ for the same system in same physical situations as in Fig.\ \ref{csfirh}. Red solid lines show the best-fitted approximating distributions $P_{\rm Poi-2xPoi}(\lambda;S)$ (\ref{pspoi2poi}) [panels (b), (c)] or $P_{\rm Poi-GUE}(\lambda_{\rm fit};S)$ (\ref{pspoigue}) [panels (e), (f)] with $\lambda_{\rm fit}$ specified for each plot. Remaining lines are same as in Fig.\ \ref{csfirh}.
}
\end{figure}

\begin{figure}
\centerline{\includegraphics[width=\linewidth]{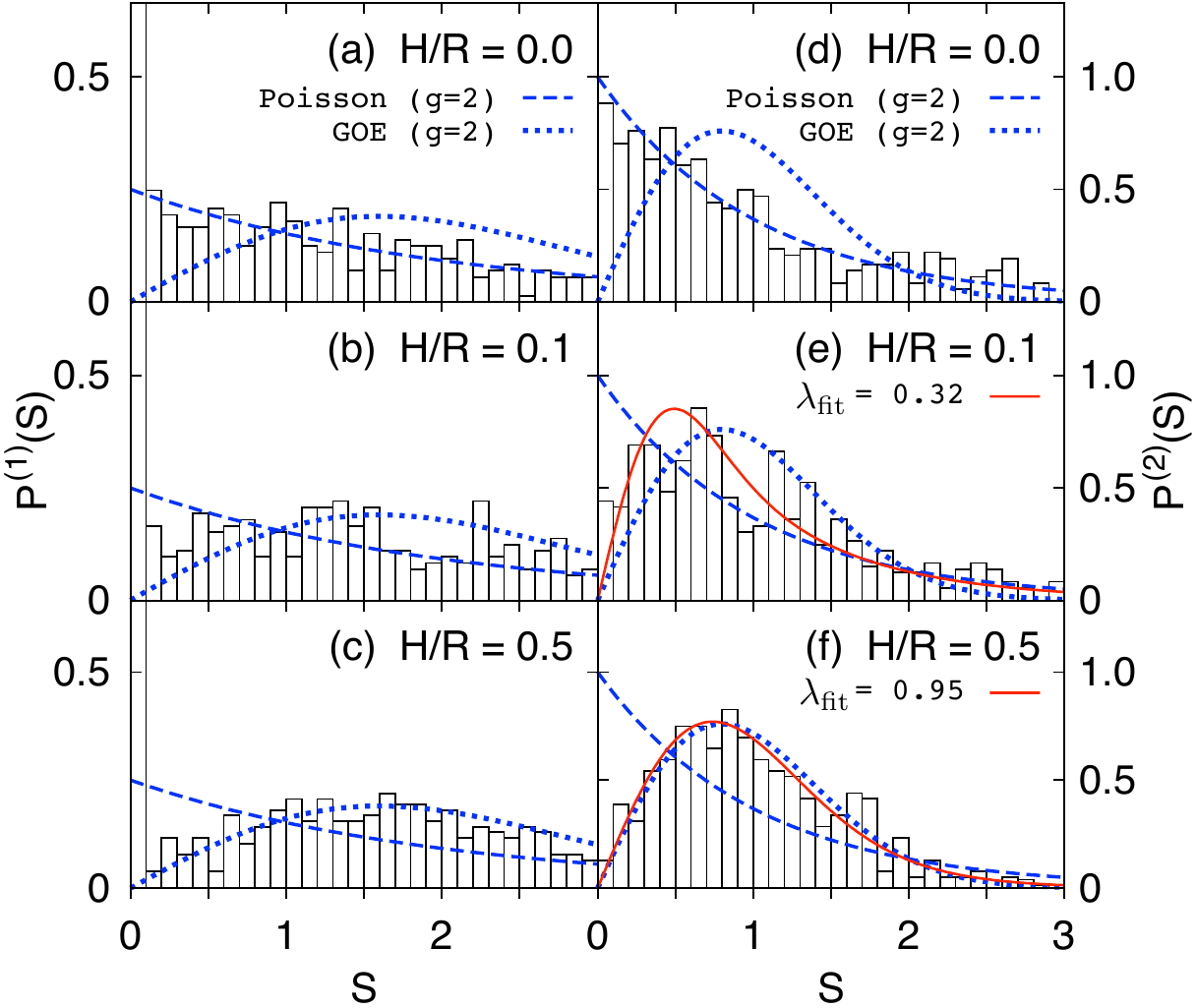}}
\caption{\label{psrhab}
  Level-spacing distributions $P^{(1)}(S)$ (a)--(c) and $P^{(2)}(S)$ (d)--(f) for the same system as in Figs.\ \ref{levrep}--\ref{psfirh} in the presence of zz-strain (with the parameter $H/R$ specified for each panel). Numerical results are shown with black solid lines. Remaining lines are for Poisson distribution with the degeneracy $g=2$ of each level [blue dashed], the Wigner surmise for GOE [blue dotted], or for the best-fitted approximating distributions $P_{\rm Poi-GOE}(\lambda_{\rm fit};S)$ (\ref{pspoigoe}) [red solid lines on panels (e), (f)] with $\lambda_{\rm fit}$ specified for each plot.
}
\end{figure}

\begin{figure}
\centerline{\includegraphics[width=\linewidth]{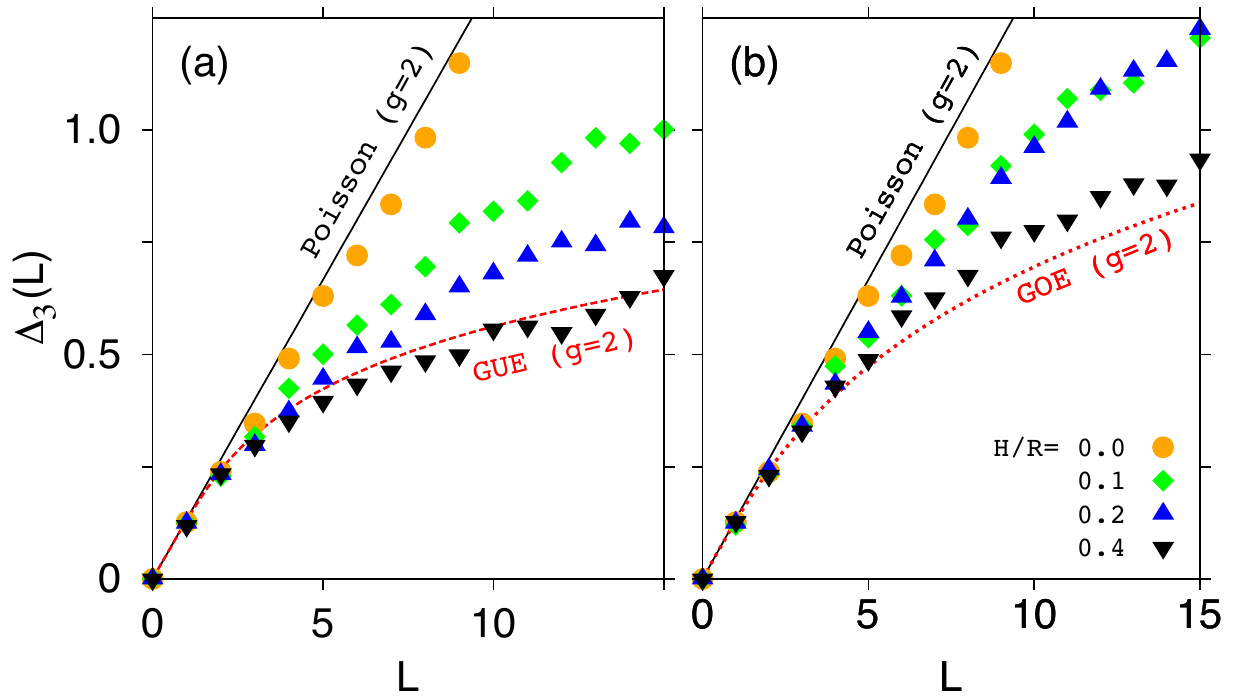}}
\caption{\label{del3th}
  Spectral rigidity $\Delta_3(L)$ for the same system as in Figs.\ \ref{levrep}--\ref{psrhab} in the presence of ac-strain (a) or zz-strain (b). Datapoints show the results obtained numerically for different values of the strain parameter $H/R$ (specified for each dataset). Lines are the theoretical expectations for the relevant ensembles of random matrices.
}
\end{figure}

The above-mentioned basic signatures of quantum chaos, accompanying strained-induced gauge fields, are further supported with spectral statistics presented in Figs.\ \ref{csfirh}--\ref{del3th}.

\subsection{Level-spacing distributions}
First, we discuss the level-spacing distributions $P^{(1)}(S)$ and their integrals $C^{(1)}(S)\equiv\int_0^SP^{(1)}(S')dS'$ (see Fig.\ \ref{csfirh}), with $P^{(k)}(S)$ ($k=1,2,\dots$) being the probability that the quantity $\langle\rho(E)\rangle(E_{n+k}\!-\!E_n)/k$ is located in the interval $(S,S\!+\!dS)$. $E_{n+k}\!-\!E_n$ is the distance between $k$-th neighbors in the level sequence $E_1\leqslant{}E_2\leqslant\dots$, and $\langle\rho(E)\rangle$ is the average density of levels in the energy interval $(E,E+dE)$, which can be approximated by $\langle\rho(E)\rangle\simeq{\pi}^{-1}{\cal A}_{\rm S}|E|/(\hbar{}v_F)^2$ for $|E|\ll{}t_0$. The numerical results, shown with black solid lines in Fig.\ \ref{csfirh}, are obtained for about $800$ energy levels $0<E_n<0.5\,t_0$ \cite{unfofoo}. The theoretical curves (blue lines) are given by
\begin{equation}
  \label{psxg12}
  P_{X,g}^{(1)}(S)=\begin{cases}
    P_X(S), & \text{if  } g=1, \\
    \frac{1}{2}\delta(S)+\frac{1}{4}P_X(S/2), & \text{if  } g=2,
  \end{cases}
\end{equation}
with $g=1,2$ the level degeneracy, and
\begin{equation}
  \label{psxwig}
  P_X(S)=\begin{cases}
    \exp(-S), & \!\!\text{for Poisson}, \\
    (\pi/2)S\exp\!\left(-\pi{}S^2\!/4\right), & \!\!\text{for GOE}, \\
    (32/\pi^2)S^2\!\exp\!\left(-4S^2\!/\pi\right), & \!\!\text{for GUE},
  \end{cases}
\end{equation}
where we have used the Wigner surmise approximating $P^{(1)}(S)$ for the relevant ensemble of random matrices \cite{Haa10}. The evolution of spectral statistics $C^{(1)}(S)$ and $P^{(1)}(S)$ with the magnetic field, illustrated in Figs.\ \ref{csfirh}(a)--(c), confirms that the system energy spectrum gradually transforms from Poissonian with the twofold valley degeneracy (manifesting itself for $\Phi\simeq{}0$) towards Poissonian without such a~degeneracy (approached for $\Phi\gtrsim\Phi_0$), showing no signatures of quantum chaos in the absence of strain fields. The evolution of same statistics with the varying ac-strain and fixed $\Phi=0$ [Figs.\ \ref{csfirh}(d)--(f)] unveils the spectral fluctuations characteristic for GUE, starting from $H/R\gtrsim{}0.1$. 

The corresponding results for the second-neighbor spacing distributions $P^{(2)}(S)$ are presented in Fig.\ \ref{psfirh}. The theoretical expectations depicted with blue lines follow from Eq.\ (\ref{psxwig}) via
\begin{equation}
  P_{X,g}^{(2)}(S)=\begin{cases}
    2\int_0^{2S}dS'P_X(2S\!-\!S')P_X(S'), & \text{if  } g=1, \\
    P_X(S), & \text{if  } g=2.
  \end{cases}
\end{equation}
[For instance, $P^{(2)}_{{\rm Poi},\,g=1}(S)=4S\exp(-2S)$.] Additionally, we have utilized additive random-matrix models of the form \cite{Ryc12,Zyc93}
\begin{equation}
  \label{hamlam}
  H(\lambda)=\frac{H_0+\lambda{}V}{\sqrt{1+\lambda^2}},
\end{equation}
allowing one to generate (for $0<\lambda<\infty$) the spacing distributions interpolating between that of two distinct ensembles of random matrices.

\begin{widetext}
Three particular choices of $H_0$ and $V$ lead to: 
\begin{itemize}

\item
A distribution interpolating between Poisson with $g=2$ (reproduced for $\lambda=0$) and Poisson with $g=1$ (reproduced for $\lambda=\infty$) [see Figs.\ \ref{psfirh}(a)--(c)]
\begin{equation}
  \label{pspoi2poi}
  P_{{\rm Poi-2xPoi}}(\lambda;S)= 
  \left[\frac{1+a(\lambda)}{1-a(\lambda)}\right]\exp(-S)
  \left\{ \exp[-Sa(\lambda)] 
    - \exp\left[-\frac{S}{a(\lambda)}\right] \right\}.
\end{equation}

\item
A distribution interpolating between Poisson and GOE [see Fig.\ \ref{psrhab}]
\begin{equation}
  \label{pspoigoe}
  P_{\rm Poi-GOE}(\lambda;S)=
  \left[\frac{u(\lambda)^2S}{\lambda}\right]
  \exp\left[{-\frac{u(\lambda)^2S^2}{4\lambda^2}}\right]
  \int_0^\infty\!{d\eta}\,\exp(-\eta^2-2\lambda\eta)
  I_0\left[\frac{\eta{u(\lambda)}S}{\lambda}\right].
\end{equation}

\item
A distribution interpolating between Poisson and GUE [see Figs.\ \ref{psfirh}(d)--(f)]
\begin{equation}
  \label{pspoigue}
  P_{\rm Poi-GUE}(\lambda;S)=
  \sqrt{\frac{2}{\pi}}\left[\frac{c(\lambda)^2S}{\lambda}\right]
  \exp\left[{-\frac{c(\lambda)^2S^2}{2\lambda^2}}\right] 
  \int_0^\infty\!\frac{d\eta}{\eta}\,\exp\left(-\lambda\eta-\frac{\eta^2}{2}\right)\sinh\left[\frac{\eta{c(\lambda)}S}{\lambda}\right].
\end{equation}

\end{itemize}
The coefficients $a(\lambda)$,  $u(\lambda)$, and $c(\lambda)$ are chosen such that $\langle{}S\rangle_X=\int_0^\infty{}SP_X(\lambda;S)dS=1$ for any value of $\lambda$ \cite{abcfoo}; $I_0(x)$ in Eq.\ (\ref{pspoigoe}) is the modified Bessel function of the first kind.
\end{widetext}

Formally, Eq.\ (\ref{pspoi2poi}) represents the exact expression for second-neighbor spacing distribution in case $H_0$ is diagonal random matrix with twofold degeneracy of each eigenvalue and $V$ is diagonal random matrix without such a~degeneracy. Similarly, Eqs.\ (\ref{pspoigoe}) and (\ref{pspoigue}) represent the exact level-spacing distributions for $2\times{}2$ random matrices of the form $H(\lambda)$ (\ref{hamlam}), with $H_0$ chosen as diagonal random matrix and $V$ being a~member of GOE or GUE (respectively). It was found numerically, however, that such distributions approximate, with a~suprising accuracy, the actual level-spacings distributions of large random matrices \cite{Zyc93} as well as distributions obtained for various dynamic systems undergoing transitions between symmetry classes \cite{Haa10,Ber86,Zyc93}.

The distribution $P_{\rm Poi-2xPoi}(\lambda;S)$ (\ref{pspoi2poi}), with the best-fitted parameter $\lambda=\lambda_{\rm fit}$, is capable of reproducing the evolution of $P^{(2)}(S)$ when the valley degeneracy is being split by external magnetic field in the absence of strain fields [see red lines in Figs.\ \ref{psfirh}(b) and \ref{psfirh}(c)]. Similarly, the distribution $P_{\rm Poi-GUE}(\lambda_{\rm fit};S)$ (\ref{pspoigue}) is capable of reproducing $P^{(2)}(S)$ when the system undergoes transition to quantum chaos induced by ac-strain [see red lines in Figs.\ \ref{psfirh}(e) and \ref{psfirh}(f)]. In the latter case, $P^{(2)}(S)$ exhibits transition Poisson-GUE analogous to the transition demonstrated numerically in Ref.\ \cite{Ryc12} for the similar system with smooth potential disorder. 

The evolution of spacing distributions $P^{(1)}(S)$ and $P^{(2)}(S)$ with zz-strain is illustrated in Fig.\ \ref{psrhab}. This time, transition to quantum chaos manifests itself by a~systematic crossover of the actual spectral statistics obtained numerically (black solid lines) between the theoretical predictions for Poisson and GOE (blue lines) with the twofold valley degeneracy preserved. Also, the approximating distribution $P_{\rm Poi-GOE}(\lambda_{\rm fit};S)$ (\ref{pspoigoe}) is capable of reproducing $P^{(2)}(S)$ during the transition Poisson-GOE [see red lines in Figs.\ \ref{psrhab}(e) and \ref{psrhab}(f)].

\subsection{Spectral rigidity}
The fluctuations of more distant spacings between energy levels of quantum system can be described in a~compact way by the spectral rigidity \cite{Dys63}
\begin{equation}
  \label{del3def}
  \Delta_3(L)=\frac{1}{L}\Big<
  \underset{\scriptsize (a,b)}{\mbox{Min}}
  \int_{-L/2}^{L/2}\!dx\left[{\cal N}(x_0\!+\!x)-ax-b\right]^2
  \Big>,
\end{equation}
where $x\equiv\langle{\cal N}(E)\rangle$ and ${\cal N}(E)$ denotes the number of energy levels such that $0<E_n\leqslant{}E$. Theoretical expectations $\Delta_3^{\rm (Poi)}(L)$, $\Delta_3^{\rm (GOE)}(L)$, and $\Delta_3^{\rm (GUE)}(L)$ are given explicitly in Refs.\ \cite{Ryc12,Dys63}. For the case of $g$-fold degeneracy of each energy level one finds immediately from Eq.\ (\ref{del3def}) that the corresponding theoretical expression needs to be rescaled via $\Delta_3^{(g,X)}(L)=g^2\Delta_3^{(X)}(L/g)$. 

The numerical values of $\Delta_3(L)$ for our model system are shown in Fig.\ \ref{del3th} together with theoretical expectations for the relevant ensembles of random matrices. Similarly as for the statistics $P^{(1,2)}(S)$ discussed above, the evolution of $\Delta_3(L)$ with ac-strain clearly exhibits transition Poisson-GUE [see Fig.\ \ref{del3th}(a)] whereas the evolution of $\Delta_3(L)$ with zz-strain exhibits transition Poisson-GOE [see Fig.\ \ref{del3th}(b)], with the valley degeneracy ($g=2$) preserved in the absence of magnetic field.

\begin{figure}
\centerline{\includegraphics[width=\linewidth]{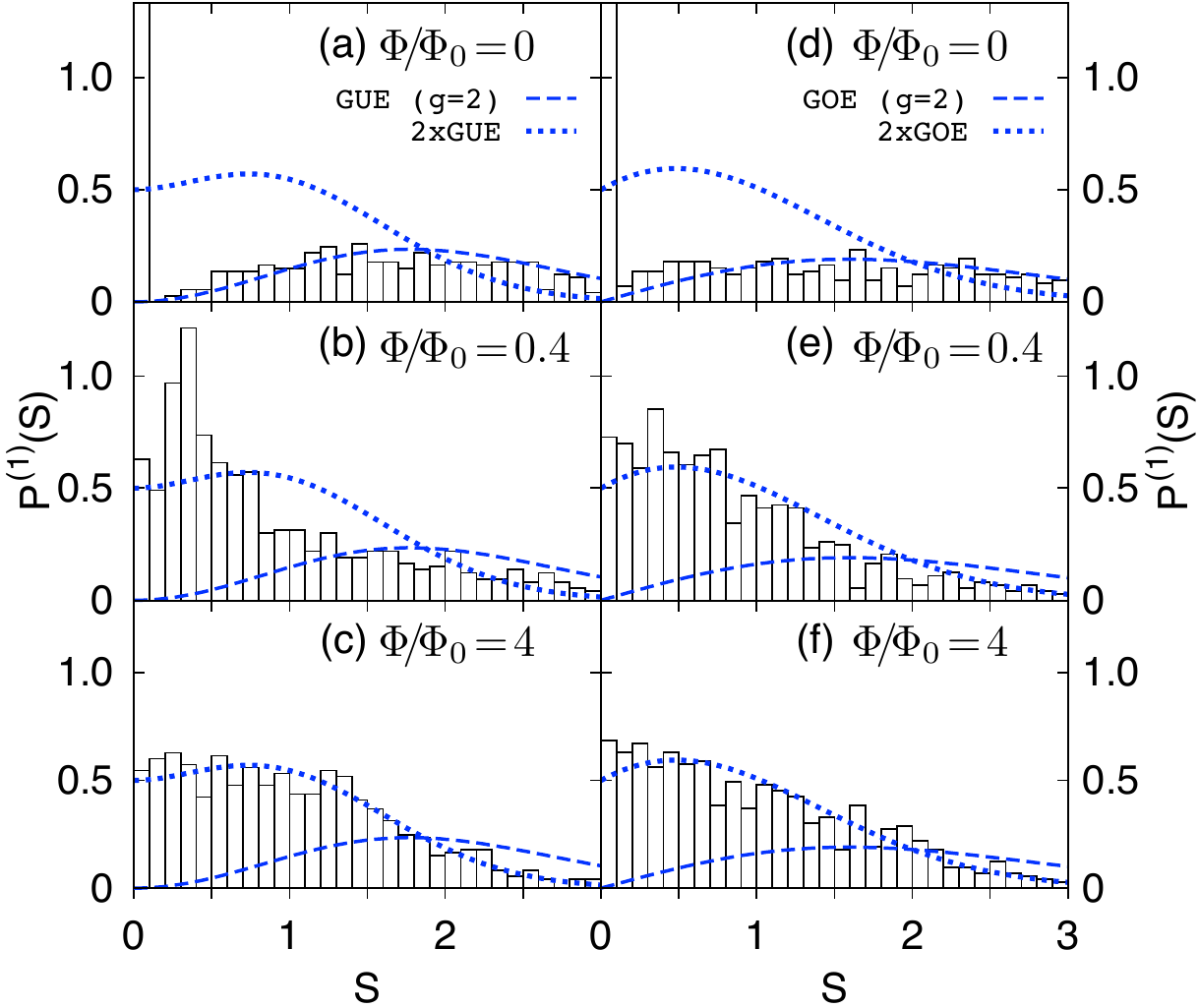}}
\caption{\label{psfirhab}
  Level-spacing distributions $P^{(1)}(S)$ for the same system as in Figs.\ \ref{levrep}--\ref{del3th} in the presence of uniform magnetic field (with the flux $\Phi$ varied between the panels) and persistent ac-strain [panels  (a)--(c)] or zz-strain [panels (d)--(f)], with the strain parameter fixed at $H/R=0.2$ in both cases. Numerical results are shown with black solid lines. Remaining lines are for GUE with the degeneracy $g=2$ of each level [blue dashed], or for two independent GUEs [blue dotted].
}
\end{figure}

\begin{figure}
\centerline{\includegraphics[width=\linewidth]{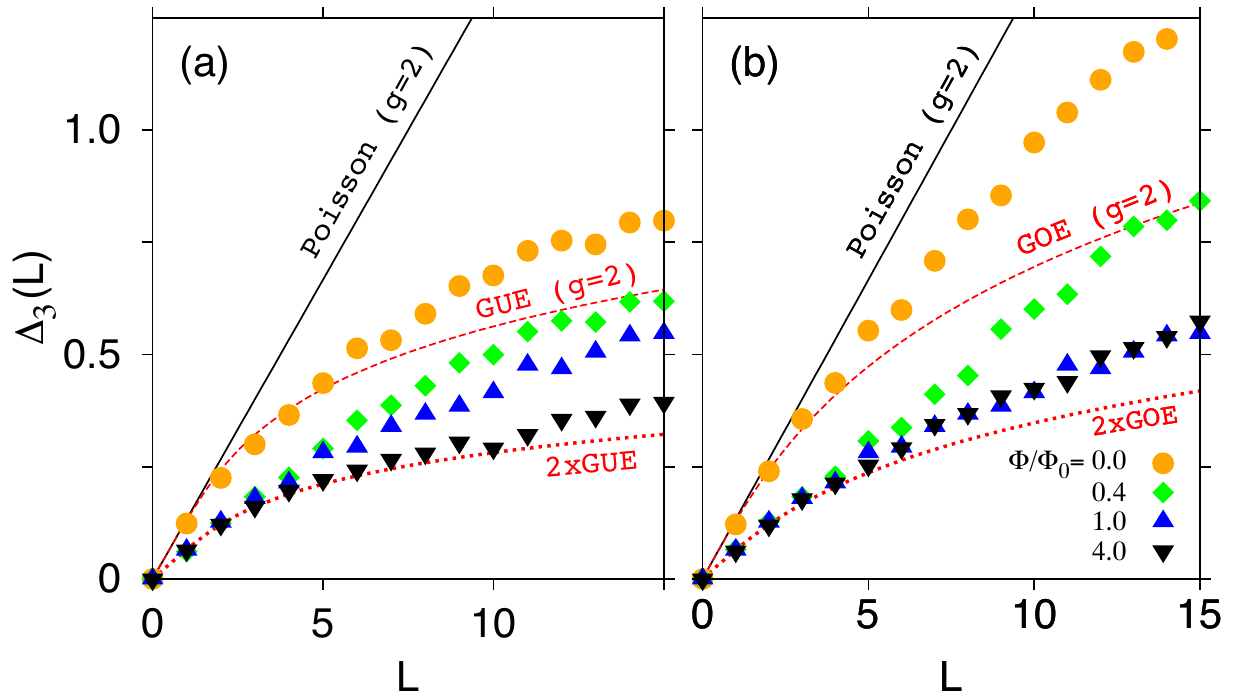}}
\caption{\label{del3firh}
  Same as Fig.\ \ref{del3th} but in the presence uniform magnetic field ($\Phi/\Phi_0$ is specified for each dataset) and persistent ac-strain (a) or zz-strain (b); $H/R=0.2$ in both cases.
}
\end{figure}

\subsection{Systems with persistent strains in external magnetic fields}
For a~sake of completeness, we consider now the systems of Fig.\ \ref{strageo} with the strain parameter fixed at $H/R=0.2$ (the cases of ac-strain and zz-strain are studied separately) placed in uniform magnetic fields varying in the range $0\leqslant{}\Phi/\Phi_0\leqslant{}4$ [see Figs.\ \ref{psfirhab} and \ref{del3firh}]. Although the discussion is still limited to the deformations given by Eqs.\ (\ref{acstra}) and (\ref{zzstra}), the universal nature of spectral fluctuations in the limit of quantum chaos allows us to believe that the effects which we demonstrate numerically in this subsection may be observable for graphene nanoflakes with persistent (for instance, substrate-induced) strains, such as studied experimentally in Refs.\ \cite{Lev10,Ham11}. 

The numerical results for level-spacing distributions $P^{(1)}(S)$ are shown in Fig.\ \ref{psfirhab} (black solid lines). The theoretical predictions for GOE and GUE (with the valley degeneracy) are given by Eqs.\ (\ref{psxg12},\ref{psxwig}) and drawn with blue dashed lines. The predictions for level sequences following from two statistically-independent GOEs or GUEs (blue dotted lines) are given by expressions derived by Robnik and Berry \cite{Rob86}
\begin{equation}
  \label{psd2es}
  P^{(1)}_{2\times{}X}(S) = \frac{d^2}{dS^2}\left[ E_{X}(S/2) \right]^2, 
\end{equation}
with 
\begin{equation}
  \label{esxwig}
  E_X(S)=\begin{cases}
    1-\mbox{erf}\left(S\sqrt{\pi}/2\right), & \text{for GOE}, \\
    \exp\left(-4S^2/\pi\right) \\
    \ \ \ \ \ \ -S+S\,\mbox{erf}\left(2S/\sqrt{\pi}\right), & \text{for GUE},
  \end{cases}
\end{equation}
being the probability that interval S contains no energy level of a~single sequence following GOE or GUE. The error function $\mbox{erf}(x)=(2/\sqrt{\pi})\int_0^x\exp(-t^2)dt$. Hereinafter, we suppose identical densities of energy levels in both components of a~combined sequence.

The datasets presented in Figs.\ \ref{psfirhab}(a)--(c) illustrates a~systematic crossover of the actual spectral statistics $P^{(1)}(S)$ between the theoretical predictions for GUE with the valley degeneracy $g=2$ and two independent GUEs driven by external magnetic field in the presence of persistent ac-strain. Similarly, a~field-driven crossover between GOE (with $g=2$) and two independent GOEs is clearly visible in Figs.\ \ref{psfirhab}(d)--(f) in the situation with persistent zz-strain. Additionally, both crossovers between the relevant ensembles of random matrices are visualized with the spectral rigidity $\Delta_3(L)$; see respectively Figs.\ \ref{del3firh}(a) and \ref{del3firh}(b) for the cases of persistent ac-strain and persistent zz-strain. [Notice that theoretical expectations for two independent GOEs or GUEs are $\Delta_3^{(2\times{}X)}(L)=2\Delta_3^{(X)}(L/2)$.] Remarkably, either $P^{(1)}(S)$ or $\Delta_3(L)$ approaches the theoretical predictions for two independent GOEs (or GUEs) when $\Phi\gtrsim{}\Phi_0$, and the (approximate) valley degeneracy no longer affects the electronic spectrum of the system. 

It is worth to stress that, apart from lifting up the valley degeneracy, weak magnetic fields do not alter the symmetry classes of chaotic Dirac systems we consider here. In the absence of intervalley scattering, turning on the magnetic field simply transforms a~system into a~pair of two independent chaotic systems (one at each valley) each of which is showing the same symmetry class as the original system at zero field: Namely, the unitary if strain fields break STRS (the case of ac-strain of a~generic nature) or the orthogonal if a~mirror symmetry leads to false STRS breaking (the special case of zz-strain). Chaotic graphene systems with strong intervalley scattering, earlier considered in Refs.\ \cite{Wur09,Lib09}, show standard transition GOE-GUE, exhibiting another striking difference between Dirac billiards and generic graphene flakes (with irregular edges). Nevertheless, the former still can be modelled effectively within particular graphene nanoflakes, with terminal atoms belonging predominantly to one sublattice.

\begin{figure}
\centerline{\includegraphics[width=0.9\linewidth]{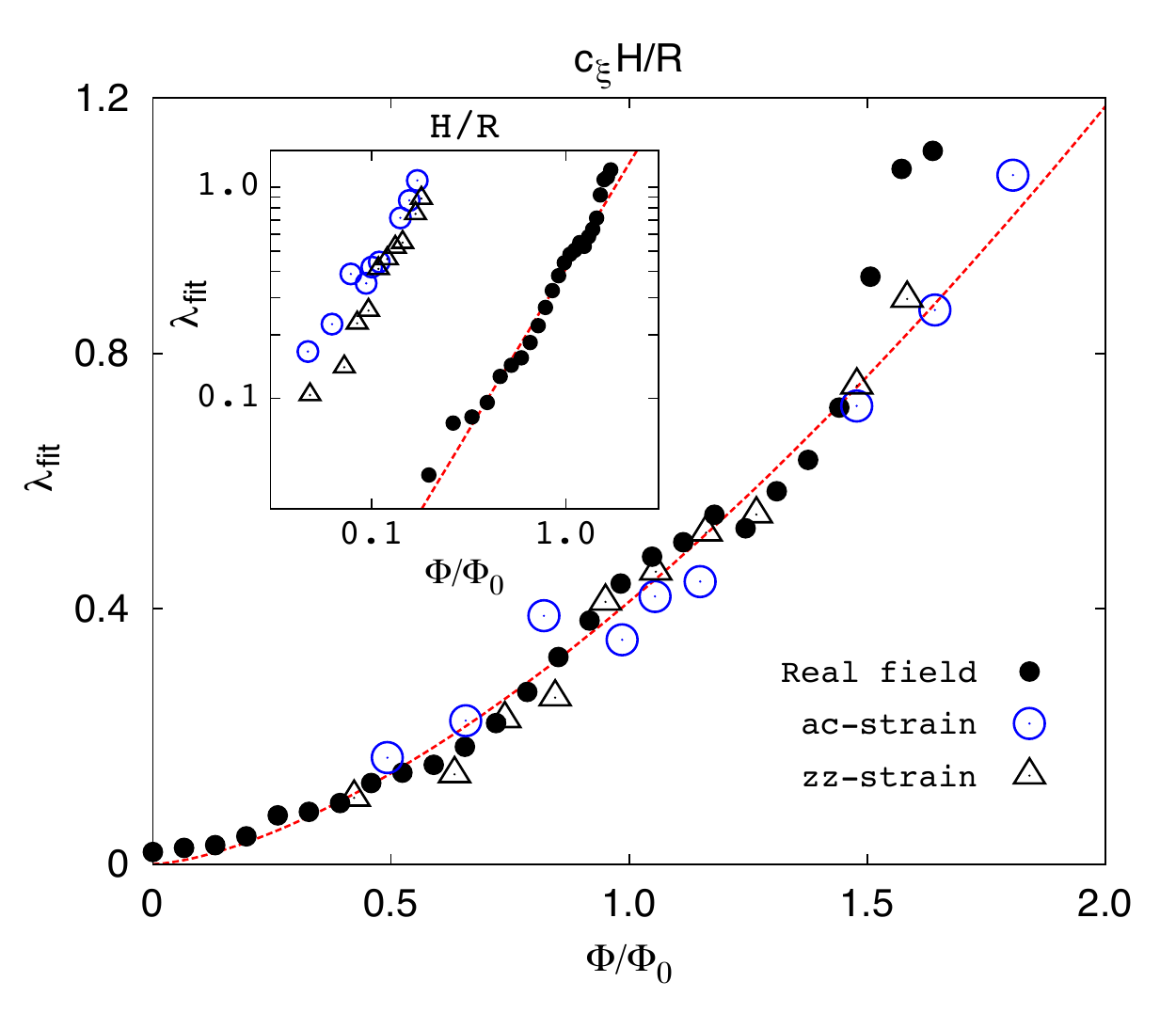}}
\caption{\label{lbfirh}
  Least-squares fitted parameters $\lambda_{\rm fit}$ for transitions between random ensembles demonstrated numerically in Sec.\ \ref{specsta} [see Eqs.\ (\ref{pspoi2poi})--(\ref{pspoigue})], as functions of the magnetic flux $\Phi$ [solid symbols], or effective fluxes $\Phi_{{\rm eff},\xi}$ [open symbols] defined by Eqs.\ (\ref{phieffxi}) and (\ref{cxivals}). Dashed line represent the best-fitted power-low relation given by Eq.\ (\ref{lbfitsim}). Inset shows the data for strained nanosystems directly as functions of the strain parameter $H/R$, compared with the data obtained for real magnetic fields, in the log-log scale. 
}
\end{figure}

\section{ Relation between strain fields and real magnetic fields \label{reremag} }

We supplement our numerical study of quantum chaos in strained graphene nanoflakes by comparing, in a~quantitative manner, the effects of geometric deformations given by Eqs.\ (\ref{acstra}) and (\ref{zzstra}) with the effects of real magnetic fields on the spacing distribution $P^{(2)}(S)$.

As discussed in Sec.\ \ref{specsta}, a~significant pseudomagnetic field \emph{per se}, $B_{\rm s}\propto{}H/R$ (\ref{bsprop}), appears for zz-strain only. However, as the actual spacing distributions $P^{(2)}(S)$ can be rationalized by $P_X(\lambda_{\rm fit},S)$ given (respectively) by Eqs.\ (\ref{pspoi2poi}), (\ref{pspoigoe}), or (\ref{pspoigue}) for the cases of real magnetic field, zz-strain, or ac-strain, the numerical comparison of the best-fitted parameters $\lambda_{\rm fit}$ (provided in Fig.\ \ref{lbfirh}) allows one to define the effective fluxes 
\begin{equation}
  \label{phieffxi}
  \frac{\Phi_{{\rm eff},\xi}}{\Phi_0}=c_\xi\frac{H}{R}\ \ \ \ 
  (\text{with  }\xi={\rm ac, zz})
\end{equation}
for either the cases of ac-strain and zz-strain. The coefficients $c_\xi$ are adjusted such that $\lambda_{\rm fit}$ for strained systems [open symbols in Fig.\ \ref{lbfirh}] follow the approximating power-law relation [red dashed lines] found for the case of real magnetic field 
\begin{equation}
  \label{lbfitsim}
  \lambda_{\rm fit}\simeq 0.41(1)\times\left[ \Phi/\Phi_0 \right]^{1.53(7)},
\end{equation}
with the numerical values of parameters obtained via least-squares fitting (the standard deviation of a last digit are specified by numbers in parenthesis). This leads to 
\begin{equation}
\label{cxivals}
  c_{\rm ac}=10.5(2)\ \ \ \ \text{and}\ \ \ \ c_{\rm zz}=8.8(1).
\end{equation}
We notice, that the dimensionless parameter $c_{\rm zz}$ given by Eq.\ (\ref{cxivals}) corresponds to $c_{\rm zz}\Phi_0/{\cal A}_{\rm S}\simeq{}43\,$T, what is numerically very close to the value reported by the second paper of Ref.\ \cite{Gui10b} for the limit of Landau quantization. Therefore, the relation between uniform magnetic and strain-induced pseudomagnetic fields, following from the statistical description of the evolution of energy levels in weak fields by means of additive random-matrix models, appears to be consistent with the corresponding relation arising from transport properties in strong fields.

\section{ Conclusions \label{conclu} }
We have discussed the selected spectral statistics of triangular graphene nanoflakes with zigzag edges in the presence of strain-induced gauge fields. Such systems may show the spectral fluctuations following GUE of random matrices, providing the link between chaotic graphene flakes \cite{Ryc12,Wur09,Lib09} and Dirac billiards for massless spin-$1/2$ fermions \cite{Ber87,XNi12}.

In the absence of disorder, strain fields associated with moderate in-plain deformations drive a highly-symmetric nanosystem into chaotic regime. Moreover, our results show, that the system symmetry class is related to the particular arrangement the deformation: In a~generic case of the deformation breaking all geometric symmetries the unitary symmetry class is observed, as the strain field also breaks the effective (symplectic) time reversal symmetry (STRS) in a~single valley. To the contrary, if a~single mirror symmetry is preserved, we have only the false STRS breaking leading to the orthogonal symmetry class. (Such a~physical situation has no analogue in graphene nanoflakes with substrate-induced disorder \cite{Ryc12} or irregular edges \cite{Wur09,Lib09}, where all geometric symmetries are naturally broken.)  It is worth to stress, that although particular strain arrangements studied here are different from those in existing experiments \cite{Boo08,Bun08,Bao09,Hua09,Moh09,Lev10}, the nature of the results allows one to expect, that  the system symmetry class remains unitary (orthogonal) for an arbitrary strain arrangement leading to the true (false) STRS breaking, providing the deformations are sufficiently small that the effective Dirac theory for low-energy excitations can be applied. 

In both cases of the unitary and the orthogonal symmetry classes, spectral statistics obtained numerically follow those of the relevant ensembles of random matrices (GUE or GOE), with the twofold valley degeneracy of each energy level. When the real magnetic and strain fields are applied simultaneously, the valley degeneracy no longer applies, and our system displays spectral fluctuations following two statistically-independent GUEs (or GOEs), one per each valley.

The evolution of spectral statistics with weak strain fields, exhibiting the transition to quantum chaos, is rationalized using additive random-matrix models. The functional relations between the best-fitted model parameters and the deformation strength allow one to compare the effects of strain fields with the effects of real magnetic fields in a~quantitative manner. The resulting numerical relation between uniform magnetic and strain-induced pseudomagnetic fields stays in agreement with the similar relation, earlier obtained for strong fields leading to the appearance of Landau quantization (see the second paper of Ref.\ \cite{Gui10b}).

Although our work was primarily motivated by the fabrication of regular graphene nanoflakes \cite{Ham11}, it may also be possible to observe the effects which we describe in artificial graphenes, such as the arrays of GaAs/AlGaAs quantum wells \cite{Sin11}, or in the recently discussed acoustic analog of graphene \cite{Tor12}.

\section*{Acknowledgments}
I thank to Ari Harju for the correspondence.
The work was supported by the National Science Centre of Poland (NCN) via Grant No.\ N--N202--031440, and partly by Foundation for Polish Science (FNP) under the program TEAM {\em ``Correlations and coherence in quantum materials and structures (CCQM)''}. Computations were partly performed using the PL-Grid infrastructure.


\begin{thebibliography}{}
\bibitem{valefoo}
In particular, real magnetic field break TRS and has the same sign in the two valleys, whereas the strain-induced gauge field preserves TRS and has opposite signs in the two valleys, see Ref.\ \cite{Voz10}.
\bibitem{dilafoo}
Hereinafter, we neglect the effects of scalar potential induced by lattice dilatation, see Ref.\ \cite{Gui10b}
\bibitem{edgefoo}
In a closed nanosystem, symmetries associated with ${\cal T}_0$ and ${\cal T}_x$ become relevant if the intervalley scattering due to the boundaries is insignificant, see Refs.\ \cite{Ryc12,Wur11}.
\bibitem{delnfoo}
For a~given strain-induced pseudomagnetic field $B_{\rm s}$, Landau quantization is observable when the average number of quantum states per one Landau level $\langle\Delta{\cal N}\rangle={\cal A}_S/(\pi{}l_s^2)\gg{}1$, with $l_s\equiv{}\sqrt{\hbar/(e|B_{\rm s}|)}$.
\bibitem{unfofoo}
We use the empirical density of states in a form
$
  \rho_{\rm eff}(E)=\rho_0+{\pi}^{-1}{\cal A}_{\rm eff}|E|/(\hbar{}v_F)^2,
$
with least-squares fitted parameters $\rho_0$ and ${\cal A}_{\rm eff}$, 
and find that it provides a~resonable approximation of $\langle{}\rho(E)\rangle$ for the whole energy range of $0<E<0.5\,t_0$ even for the strain parameters as high as $H/R\simeq{}0.5$.
Remaining details of calculating level-spacing distributions $P^{(k)}(S)$ are same as in Ref.\ \cite{Ryc12}.
\end{thebibliography}
\end{document}